\documentclass[aps,prl,showpacs,superscriptaddress,11pt]{revtex4}%
\usepackage{amsmath,amssymb,paralist}

\begin{document}

\title{Proofs of nonlocality without inequalities revisited}

\author{GianCarlo \surname{Ghirardi}}\email{ghirardi@ts.infn.it}%
\affiliation{Department of Theoretical Physics of the University of
  Trieste, Italy}%
\affiliation{Istituto Nazionale di Fisica Nucleare, Sezione di Trieste,
  Italy}%
\affiliation{International Centre for Theoretical Physics ``Abdus Salam,''
  Trieste, Italy}%

\author{Luca \surname{Marinatto}}\email{marinatto@ts.infn.it}%
\affiliation{Department of Theoretical Physics of the University of
  Trieste, Italy}%
\affiliation{Istituto Nazionale di Fisica Nucleare, Sezione di Trieste,
  Italy}%

\date{\today}

\begin{abstract}
  We discuss critically the so-called nonlocality without inequalities proofs for bipartite quantum states,
  we generalize them and we analyze their relation with the Clauser-Horne inequality.
\end{abstract}


\maketitle


\section{Introduction}

Hardy's nonlocality without inequalities proof~\cite{hardy} has been referred to as the {\em ``best version of
the Bell theorem"}~\cite{mermin}. It demonstrates that, given almost any entangled bipartite pure state of two
spin-$1/2$ particles, appropriately chosen joint-probability distributions exist which cannot be accounted for
by a local hidden variable model in which the outcomes of single-particle measurements are predetermined by the
knowledge of the hidden variables. More specifically, given any entangled but not maximally entangled state
$\psi\in {\mathbb{C}}^{2} \otimes {\mathbb{C}}^{2}$, Hardy exhibits four appropriately chosen spin-observables
dependent on the state $\psi$, which we will denote as $\left\{ X_{i}, Y_{i}\right\}$ ($i=1,2$ being the
particle index) such that, according to quantum mechanics, the following joint-probabilities hold
\begin{eqnarray}
 \label{eq1.1}
 P_{\psi}(X_{1}=+1, X_{2}=+1) &= & 0,\\
 \label{eq1.2}
 P_{\psi}(Y_{1}=+1, X_{2}=-1) &= & 0,\\
 \label{eq1.3}
 P_{\psi}(X_{1}=-1, Y_{2}=+1) &= & 0, \\
 \label{eq1.4}
 P_{\psi}(Y_{1}=+1, Y_{2}=+1) &= & q_4.
\end{eqnarray}
Then, by resorting to (counterfactual) logical reasonings Hardy's argument~\cite{hardy} succeeds in proving that
a local model, reproducing the above correlations for the considered state $\psi$, cannot exist whenever the
probability $q_{4}$ of Eq.~(\ref{eq1.4}) does not vanish. Such an argument has been commonly considered not to
make any reference to Bell-like inequalities~\cite{bell,ch} and its experimental realization~\cite{demartini},
in which one tests that the joint probabilities of Eqs.~(\ref{eq1.1})-(\ref{eq1.3}) are strictly null while the
one of Eq.~(\ref{eq1.4}) is not, provides evidence of nonlocality. However, contrary to this widespread opinion,
Hardy's nonlocality condition is explicitly expressed in terms of an inequality (trivial as it might seem, but
nonetheless an inequality), since testing that $q_4$ is not null amounts to test that it is strictly positive,
that is, $q_{4}>0$, with the supplementary constraints that $P_{\psi}(X_{1}=+1, X_{2}=+1)=P_{\psi}(Y_{1}=+1,
X_{2}=-1)= P_{\psi}(X_{1}=-1, Y_{2}=+1)=0$.

Subsequently, in the literature generalizations~\cite{kunkri} of the Hardy's argument appeared aiming at
softening some of the correlations requested by the original argument. In fact, the authors of
Ref.~\cite{kunkri} have considered another set of correlations
\begin{eqnarray}
 \label{eq2.1}
 P_{\psi}(X_{1}=+1, X_{2}=+1) &= & q_1,\\
 \label{eq2.2}
 P_{\psi}(Y_{1}=+1, X_{2}=-1) &= & 0,\\
 \label{eq2.3}
 P_{\psi}(X_{1}=-1, Y_{2}=+1) &= & 0, \\
 \label{eq2.4}
 P_{\psi}(Y_{1}=+1, Y_{2}=+1) &= & q_4.
\end{eqnarray}
and proved that a local hidden variable model reproducing such correlations, given a two spin-$1/2$ state $\psi$
and four spin-observables $\left\{ X_{i},Y_{i}\right\}$, cannot exist whenever the condition $q_1-q_4<0$ is
satisfied. Such an inequality, which is referred to as a nonlocality condition, reduces to Hardy's nonlocality
condition (that is, $q_4>0$) in the particular case $q_{1}=0$. Once again, such a nonlocality argument is
explicitly expressed through an inequality involving appropriately chosen probabilities. Now, one might wonder
if a relation exists between the Hardy-like nonlocality conditions, as those mentioned in Refs.~\cite{hardy}
and~\cite{kunkri}, and Bell-like inequalities~\cite{bell,ch}. The aim of this paper is precisely that of showing
clearly how the just mentioned relations are nothing but particular instances of a violated Clauser-Horne
inequality~\cite{ch}. Since this is the case, as we are going to prove, the supposed absence of Bell-like
inequalities in Hardy's arguments~\cite{hardy,kunkri,refine} (whence the name given to them, that is, {\em
nonlocality without inequalities proofs}) is only seeming: Hardy-like nonlocality conditions can be obtained by
plugging appropriate joint-probabilities into the logical negation of the Clauser-Horne inequality.


\section{A generalized Hardy's nonlocality argument}

Before addressing the issue of how Bell-like inequalities (or, more specifically, the Clauser-Horne
inequality~\cite{ch}) might indeed be directly used to derive the alleged nonlocality without inequalities
proofs of Refs.~\cite{hardy,kunkri}, we want to exhibit a generalized nonlocality argument which comprises as
its particular instances all the aforementioned proofs. We will develop such an argument by following the idea
underlying the Hardy-like proofs~\cite{hardy,kunkri} but using the methods (and extending the results) presented
in~\cite{gm}, where counterfactual reasonings are naturally replaced by simple set theoretic manipulations. In
fact, as we will see, given some joint probability distributions, like those of Eqs.~(\ref{eq1.1}-\ref{eq1.4}),
one can always associate to them appropriate subsets of the set of the hidden variables.

To start with, let us consider four arbitrary subsets $A,B,C,$ and $D$ of a set $\Lambda$ and a probability
measure $\mu$ over $\Lambda$ (that is, a real-valued mapping defined on the power set of $\Lambda$ such that (i)
$\mu[\emptyset]=0$ and $\mu[\Lambda]=1$,(ii) $\mu[X]\leq \mu[Y]$ for any $X\subseteq Y$, (iii) $\mu [
\bigcup_{i=1}^{\infty} X_{i}] = \sum_{i=1}^{\infty}\mu[X_{i}]$ for pairwise disjoint subsets $X_{i}\subseteq
\Lambda$). Then the following result holds true:
\begin{equation}
\label{eq1}
 \mu[A\cap B] +\mu [C] - \mu[B\cap C] +\mu[D] -\mu[A\cap D] -\mu[C\cap D]\in [0,1].
\end{equation}
{\em Proof.} We start by proving the lower bound of Eq.~(\ref{eq1}). Since $A\cap D\subseteq D$ and $B\cap
C\subseteq C$, we have that $(A\cap D)\cup (B\cap C)\subseteq C\cup D$. Then the following inequality holds
\begin{equation}
\label{eq2}
 \mu[A\cap D] +\mu[B\cap C]\leq \mu[C\cup D]+\mu[A\cap B\cap C\cap D],
\end{equation}
where use has been made of property (ii) of the measure $\mu$ and the fact that $\mu[X\cup
Y]=\mu[X]+\mu[Y]-\mu[X\cap Y]$ holds for any subsets $X,Y\subseteq \Lambda$. Considering the left-hand-side of
Eq.~(\ref{eq1}) we then have:
\begin{eqnarray}
\label{eq3} \mu[A\cap B] +\mu [C] - \mu[B\cap C] +\mu[D] -\mu[A\cap D] -\mu[C\cap D] & = &\nonumber \\
 \mu[A\cap B] -\mu[B\cap C] -\mu[A\cap D] +\mu[C\cup D] & \geq &\nonumber\\
\mu[A\cap B]-\mu[A\cap B\cap C\cap D] & \geq & 0,
\end{eqnarray}
where the equality makes use of the fact that $\mu[C]+\mu[D]-\mu[C\cap D]=\mu[C\cup D]$, the first inequality
descends from Eq.~(\ref{eq2}), and the second one follows
 from the fact that  $A\cap
B\cap C\cap D\subseteq A\cap B$.

To derive the upper bound of Eq.~(\ref{eq1}) we proceed as follows. Since, denoting as $\bar{Z}= Z-\Lambda$ the
complement of $Z$ within $\Lambda$, one has $C\subseteq \bar{B}\cup(B\cap C)$ and $D\subseteq \bar{A}\cup(A\cap
D)$, we have that $C\cup D\subseteq \bar{A}\cup \bar{B}\cup (A\cap D)\cup(B\cap C)$. Thus we have proved the
following inequality
\begin{equation}
\label{eq4}
 \mu[C\cup D] \leq \mu[\bar{A}\cup \bar{B}] + \mu[A\cap D]+\mu[B\cap C].
\end{equation}
Considering once again the left-hand-side of Eq.~(\ref{eq1}) we obtain:
\begin{eqnarray}
\label{eq5} \mu[A\cap B] +\mu [C] - \mu[B\cap C] +\mu[D] -\mu[A\cap D] -\mu[C\cap D] & = &\nonumber \\
 \mu[A\cap B] -\mu[B\cap C] -\mu[A\cap D] +\mu[C\cup D] & \leq &\nonumber\\
\mu[A\cap B]+\mu[\bar{A}\cup \bar{B}] & = & 1,
\end{eqnarray}
where the inequality descends from Eq.~(\ref{eq4}), and the final equality is implied by the relation
$\mu[\bar{A}\cup \bar{B}]=\mu[\Lambda-(A\cap B)] =1-\mu[A\cap B]$. $\hfill\blacksquare$

Let us now show how the single relation of Eq.~(\ref{eq1}), which is valid for any quadruple of arbitrary
subsets $A,B,C$, and $D$ can be used to yield a proof of nonlocality which generalizes in a natural way the
Hardy-like arguments~\cite{hardy,kunkri}. To this end, consider a statistical operator $\sigma$, not necessarily
associated to a pure state, acting on the Hilbert space ${\mathbb{C}}^{2} \otimes {\mathbb{C}}^{2}$ and four
arbitrary spin-observables $\left\{X_i,Y_i\right\}$ ($i=1,2$ denoting as before the particle index), such that
the following joint probability distributions hold
\begin{eqnarray}
 \label{eq6.1}
 P_{\sigma}(X_{1}=+1, X_{2}=+1) &= & q_1,\\
 \label{eq6.2}
 P_{\sigma}(Y_{1}=+1, X_{2}=-1) &= & q_2,\\
 \label{eq6.3}
 P_{\sigma}(X_{1}=-1, Y_{2}=+1) &= & q_3, \\
 \label{eq6.4}
 P_{\sigma}(Y_{1}=+1, Y_{2}=+1) &= & q_4.
\end{eqnarray}
where $q_{i}\in[0,1]$. The issue whether or not there exists a local and deterministic hidden variable model for
$\sigma$ which might account for Eqs.~(\ref{eq6.1})-(\ref{eq6.4}), depends on the (mutual) values of
$\left\{q_i\right\}$. To prove this, we begin by recalling that one denotes as a local and deterministic hidden
variable model~\cite{fine} any conceivable theory where the measurement outcomes $m,n$ of arbitrary single
particle observables $M_1,N_2$ are predetermined given the (hidden) variables $\lambda\in \Lambda$, $\Lambda$
being a set, and where the quantum mechanical joint probabilities $P_\sigma$ are obtained by averaging the
single particle probabilities $P_{\lambda}(M_1=m)$ and $P_{\lambda}(N_2=n)$ for the indicated outcomes, over the
(normalized to unity) positive distribution $\rho(\lambda)$ of such variables, according to
\begin{equation}
 \label{eq7}
  P_{\sigma}(M_{1}=m,N_{2}=n) = \int_{\Lambda}\,d\lambda\,
  \rho(\lambda) P_{\lambda}(M_{1}=m)P_{\lambda}(N_{2}=n).
\end{equation}
Here, we have assumed that the measurement of the observables $M_{1}$ and $N_{2}$ refer to space-like separated
events and we have taken into account the locality assumption by replacing $P_{\lambda}(M_{1}=m,N_{2}=n)$ with
the product $P_{\lambda}(M_{1}=m)P_{\lambda}(N_{2}=n)$ in the integrand of Eq.~(\ref{eq7}). In a deterministic
hidden variable model, like the one we are considering, the probabilities $P_{\lambda}$ can attain the values
$0$ or $1$ only, and every joint probability distribution can be naturally associated to a measure of an
appropriate subset of $\Lambda$. In fact, let us define $A,B,C$, and $D$, as those subsets of $\Lambda$ where
the probabilities $P_{\lambda}(X_{1}=+1),P_{\lambda}(X_{2}=+1),P_{\lambda}(Y_{1}=+1)$ and
$P_{\lambda}(Y_{2}=+1)$ take the value $+1$, respectively --- for example, $A = \left\{ \:\lambda\in \Lambda
\:\vert \:P_{\lambda}(X_{1}=+1)=1\right\}$. We can then rewrite Eqs.~(\ref{eq6.1})-(\ref{eq6.4}) in terms of the
measures $\mu[Z]=\int_{Z}\,d\lambda\, \rho(\lambda)$ of appropriate subsets $Z \subseteq\Lambda$ as follows:
\begin{eqnarray}
 \label{eq8.1}
 \mu[A \cap B] & =& q_1,\\
 \label{eq8.2}
 \mu[C] - \mu[ B\cap C] & = & q_2,\\
\label{eq8.3}
 \mu[D]- \mu[A\cap D] & = & q_3,\\
\label{eq8.4}
 \mu[ C\cap D] & = & q_4.
\end{eqnarray}
For example, Eq.~(\ref{eq8.2}) is obtained as follows
\begin{eqnarray}
\label{eq9}
  P_{\sigma}(Y_{1}=+1,X_{2}=-1) & = &\int_{\Lambda}\,d\lambda\,
  \rho(\lambda) P_{\lambda}(Y_{1}=+1)P_{\lambda}(X_{2}=-1)\\
  & = & \int_{\Lambda}\,d\lambda\, \rho(\lambda) P_{\lambda}(Y_{1}=+1)[1-
 P_{\lambda}(X_{2}=+1)] \\
 & = & \mu[C] - \mu[ B\cap C] =q_2,
 \end{eqnarray}
where use has been made of the obvious relation $P_{\lambda}(X_{2}=-1)+P_{\lambda}(X_{2}=+1)=1$ holding for any
$\lambda\in \Lambda$. By substituting the expressions of Eqs.~(\ref{eq8.1})-(\ref{eq8.4}) into the set-theoretic
relation Eq.~(\ref{eq1}), valid for arbitrary subsets, we obtain the following:  \\
\noindent{\bf Theorem.} Given a state $\sigma$ and four observables $\left\{X_i,Y_i\right\}$, if a local and
deterministic hidden variable model exists for $\sigma$ accounting for the joint probabilities of
Eqs.~(\ref{eq6.1})-(\ref{eq6.4}), then the following relation must hold:
\begin{equation}
\label{eq10} q_1+q_2+q_3-q_4\in [0,1].
\end{equation}
Alternatively, a violation of Eq.~(\ref{eq10}) implies nonlocality (and therefore nonseparability~\cite{werner})
for the considered state $\sigma$. This argument of nonlocality, which we have derived by using a set-theoretic
approach, displays the following features:
\begin{enumerate}
\item It generalizes all known so-called {\em nonlocality without inequalities proofs}
for pure bipartite states~\cite{hardy,refine,kunkri,gm}. For example, $q_1=q_2=q_3=0$ and $q_4>0$ is exactly
Hardy's condition of nonlocality~\cite{hardy}, while $q_2=q_3=0$ and $0<q_1<q_4$ has been considered in
Ref.~\cite{kunkri}, and both conditions violate the lower bound of Eq.~(\ref{eq10}). Of course, the fact that an
inequality is used to reproduce the Hardy-like arguments of Refs.~\cite{hardy,kunkri} strengthens our claim that
such nonlocality conditions are not inequalities-free, contrary to what is asserted in the literature.

\item It trivially yields a nonlocality proof for maximally entangled states. For example, given the singlet state of two
spin-$1/2$ particles, if we choose $\left\{X_1,Y_2,Y_1,X_2\right\}$ to be spin-observables referring to four
directions lying in a plane, like, e.g., in the $x-y$ plane, and we choose their directions making the angles
$0,\pi/4,\pi/2$ and $3\pi/4$ with respect to the positive direction of the $x$ axis, we get a violation of the
upper bound of Eq.~(\ref{eq10}). On the contrary, the usual Hardy-like arguments failed to detect nonlocality of
maximally entangled states because their nonlocality conditions were too restrictive.

\item Contrary to the usual nonlocality without inequalities arguments, our argument can be used completely in general
also to highlight the nonlocal features of certain mixed states, and it is not limited to pure states, as
happens in, e.g., Hardy's proof.

\item It avoids the use of counterfactual arguments and it rests simply on trivial set-theoretic manipulations
making full use of the natural relations between measures of sets and joint probability distributions.

\item It can be easily and further generalized to cover the case of statistical operators $\sigma$ acting onto
${\mathbb{C}}^{d_1} \otimes {\mathbb{C}}^{d_2}$, with $d_1,d_2>2$.
\end{enumerate}
To prove the appropriateness of the last statement, in place of the previously considered dichotomic
observables, we define, in the enlarged Hilbert space, four observables $\left\{X_i,Y_i\right\}$ such that the
spectrum of $X_i$ is the set $\left\{-1,0+1\right\}$, while the one of $Y_i$ is only requested to contain the
value $+1$. Then, we add to the set of Eqs.~(\ref{eq6.1})-(\ref{eq6.4}) two more relations
\begin{eqnarray}
\label{eq11.1}
 P_{\sigma}(Y_{1}=+1, X_{2}=0)&= &q_5,\\
\label{eq11.2}
 P_{\sigma}(X_{1}=0, Y_{2}=+1)& = &q_6.
\end{eqnarray}
If we add Eq.~(\ref{eq11.1}) to~(\ref{eq6.2}) and Eq.~(\ref{eq11.2}) to~(\ref{eq6.3}) and suppose that a local
hidden variable model exists for $\sigma$, then Eqs.~(\ref{eq8.2}) and~(\ref{eq8.3}) are slightly modified as
follows:  $\mu[C] - \mu[ B\cap C]=q_2+q_5$ and $\mu[D]- \mu[A\cap D]=q_3+q_6$ respectively. Here, use has been
made of the relation $P_{\lambda}(X_{i}=-1)+P_{\lambda}(X_{i}=0)+P_{\lambda}(X_{i}=+1) =1$, valid for any
$\lambda\in\Lambda$. Hence, by replacing $(q_2,q_3)$ with the modified probabilities $(q_2+q_5,q_3+q_6)$,
respectively, the relation of Eq.~(\ref{eq10}) becomes
\begin{equation}\label{eq12}
 q_1 +q_2 +q_3+q_5+q_6-q_4  \in [0,1]
\end{equation}
which must be satisfied in order that a local and deterministic hidden variable model exists for $\sigma$. Once
again, violation of such a relation automatically implies the nonlocality (and the nonseparability) of the
considered statistical operator $\sigma$.


\section{Clauser-Horne inequality}

The present formulation of a generalized Hardy's nonlocality argument in terms of an inequality involving
measures of sets is particularly useful in providing clear evidence that the supposed absence of Bell-like
inequalities in those arguments is only seeming. In fact, if we express Eq.~(\ref{eq12}) explicitly in terms of
the associated quantum joint probabilities $P_{\sigma}$, and we assume that a local hidden variable model
exists, one easily sees, by simple manipulations, that it becomes equal to
\begin{eqnarray}
 \label{eq13}
& & P_{\sigma}(X_{1}=+1,X_{2}=+1)-P_{\sigma}(Y_{1}=+1,X_{2}=+1)-P_{\sigma}(X_{1}=+1,Y_{2}=+1)\nonumber\\
& & -P_{\sigma}(Y_{1}=+1,Y_{2}=+1)+P_{\sigma}(Y_{1}=+1)+P_{\sigma}(Y_{2}=+1)\in [0,1]
\end{eqnarray}
which is exactly the inequality derived by Clauser-Horne~\cite{ch}. Moreover, the same manipulations shaw that
Eq.~(\ref{eq12}) is implied by Eq.~(\ref{eq13}).

It is worth summarizing again the main point of this paper. As they have been presented in the literature, the
nonlocality without inequalities proofs for bipartite states~\cite{hardy,refine,kunkri,gm} start by considering
a set of joint probability distributions
--- which are usually particular instances of Eqs.~(\ref{eq6.1})-(\ref{eq6.4})
--- and exhibit a logical contradiction with the existence of a local model accounting for them. Now,
probability distributions in a hidden variable model are directly expressible as measures of appropriate subsets
and the contradiction with the locality assumption is achieved by exhibiting relations between such set-measures
which turn out to be violated. More precisely, evidence of nonlocality is given by a violation of
Eq.~(\ref{eq12}) --- or, in a restricted scenario, of Eq.~(\ref{eq10})
--- which has been proven to be completely equivalent to the Clauser-Horne inequality Eq.~(\ref{eq13}).

Concluding, we have shown that the existing Hardy-like {\em nonlocality without inequalities} proofs for
bipartite states, which consider particular choices of the probabilities $\left\{q_i\right\}$ in
Eqs.~(\ref{eq6.1})-(\ref{eq6.4}), are simply particular instances of violations of the Clauser-Horne inequality.
Even more, also the general proof we have worked out in this paper is based on the set-theoretic relation of
Eq.~(\ref{eq1}), and this relation is identical to the Clauser-Horne condition Eq.~(\ref{eq12}).


\thanks{Work supported in part by Istituto Nazionale di Fisica Nucleare, Sezione di Trieste, Italy.}


\end{document}